# OVERCOMING COGNITIVE DISTRACTION AND MEASUREMENT NOISE: STRATEGIES FOR HUMANS AND ENGINEERING SYSTEMS

**Mehdi Delrobaei**
Department of Mechatronics, K. N. Toosi University of Technology, Tehran, Iran, and
Department of Electrical and Computer Engineering, Western University, London, ON, Canada
delrobaei@kntu.ac.ir

**ABSTRACT**

Cognitive distraction and measurement noise are two distinct factors that significantly impact the performance of humans and engineering systems. Cognitive distraction occurs when an individual's attention is diverted from a task, while measurement noise refers to the random variation that can occur in system measurements. Although humans and engineering systems employ different methods to overcome these obstacles, the ultimate goal is to achieve optimal performance. An intriguing question arises: what are the similarities and differences between using the term "noise" in engineering and cognitive psychology? Additionally, it is worthwhile to explore whether the human brain and engineering control systems use similar or different approaches to attenuate noise. While this article does not provide a definitive answer, it emphasizes the importance of addressing this question and encourages further investigation. Furthermore, assistive cyber-physical devices such as wearable sensory-substitution navigation aids are highlighted as quintessential human-in-the-loop systems where engineering noise attenuation and human cognitive distraction mitigation must converge.

## 1 Introduction

The term *noise* has different meanings in engineering systems and cognitive psychology. In the study of human cognition, distraction may be used instead of noise to describe how external stimuli interfere with an individual's cognitive processes and affect their understanding and behavior. In engineering systems, noise refers to any unwanted or random signals that affect the quality of a transmission or signal.

This parallel concept raises the question of what similarities exist between human cognition and engineering systems in terms of attenuating and filtering out the effects of noise and distraction. Although this article does not claim to provide a definitive answer, it seeks to highlight the importance of this question and encourage further investigation.

One possible area of comparison between human cognition and engineering systems is using filters and algorithms to reduce the impact of noise and other unwanted signals. In both cases, the goal is to identify, eliminate or minimize the impact of irrelevant or distracting signals while preserving the integrity of the relevant information.

In human cognition, attentional processes [1] allow us to selectively focus on the most relevant information and filter out distractions. Similarly, in engineering systems, attentional mechanisms can be used to selectively amplify or attenuate signals based on their relevance and importance.

Attention mechanisms are a type of machine learning technique used in engineering systems. They enable the models to concentrate on specific parts of the input data while making predictions. They have diverse applications including natural language processing, computer vision, and speech recognition. Attention mechanisms are particularly useful in engineering systems as they help models identify important features in complex datasets which can enhance their accuracy and performance.

For instance, attention mechanisms can help a model recognize essential features in an image, such as the texture of different regions or the location of objects. They can also assist a model in identifying vital words in a sentence, which is beneficial for tasks such as sentiment analysis or machine translation.

## 2 Definitions

Cognitive distraction is a phenomenon where a person's attention is diverted from the task at hand, leading to decreased performance. It can arise for various reasons, such as fatigue, stress, or external distractions. Cognitive distraction can lead to errors in decision-making, reduced situational awareness, and decreased performance.

Some examples of cognitive distraction in our everyday lives are as follows.

- Driving while using a mobile phone is one of the most common examples of cognitive distraction. When drivers use mobile phones while driving, their attention is diverted from the road, leading to decreased situational awareness and an increased risk of accidents.
- When people try to perform multiple tasks simultaneously, their attention is divided, leading to decreased performance. For example, a student who tries to study while watching TV may be unable to retain the information effectively.
- When tired, our attention span decreases, leading to reduced performance. For example, a tired driver may not be able to react quickly to a sudden change in traffic.
- When under stress, our attention is diverted from the required task, which can result in decreased performance. For instance, a surgeon who is stressed may make errors during surgery.

In engineering systems, measurement noise is the random variation in the measurements of a system, which can lead to inaccurate results. Measurement noise can arise from various factors, such as sensor errors and environmental factors, affecting the system's performance by leading to approximate measurements.

Some examples of measurement noise in engineering systems are as follows.

- Thermal noise arises due to the random motion of electrons in a conductor. It can lead to inaccurate measurements in electronic circuits and affect the system's performance.
- Quantization noise arises when an analog signal is converted to a digital signal. It can lead to inaccurate measurements in digital systems and affect performance.
- Environmental noise is a type of measurement noise that arises due to external factors such as temperature, humidity, and electromagnetic interference. It can lead to inaccurate measurements in systems sensitive to environmental factors.
- System noise arises due to the internal noise sources in a system. It can lead to inaccurate measurements in systems sensitive to internal noise sources.

## 3 Selective Attention

Generally, humans can overcome cognitive distraction by using various strategies such as switching to a distraction-free environment, setting daily intentions, working on more challenging projects, and setting artificial project deadlines. These strategies help to improve focus, reduce stress, and increase productivity.

Similar to the engineering system, our brain may be bombarded with sensory noises and distractions, mainly in the form of sounds or sights. Our attention capacity is limited, so we can't pay equal attention to everything around us. Selective attention helps us allocate our mental resources effectively.

The human brain uses a variety of cognitive mechanisms to selectively attend to different stimuli. One of the most common mechanisms is known as the cocktail party effect, which allows us to focus on a single conversation in a noisy environment. Other mechanisms include (1) feature-based attention, which allows us to focus on specific features of an object, such as its color or shape, and (2) spatial attention, which allows us to focus on specific locations in our visual field.

In terms of performance improvement, selective attention can be a valuable tool. By focusing on the most important information and filtering out irrelevant details, we can improve our ability to complete complex tasks and make better decisions. This can increase productivity, better time management, and reduce stress levels.

## 4 Noise Attenuation

Engineering systems may overcome measurement noise using techniques such as proper shielding, cabling, and termination, as well as redundancy, error correction, and other algorithmic techniques that can help mitigate the effects of measurement noise. These techniques help to improve the accuracy and reliability of the measurements.

For instance, the Kalman filter is a powerful algorithm used for estimating system parameters. It excels at handling noisy or inaccurate measurements to estimate the state of a variable or another unobservable variable with greater accuracy. The Kalman filter estimates the system's state and predicts how it will change over time. It then compares this estimate with observed measurements, considering the uncertainties associated with both the expected and observed values.

As another example, in mobile robotics, one of the main challenges is developing precise parking control mechanisms. As a mobile robot moves around, it should be able to accurately identify its position and orientation to park itself. Our previous research [2] showed that by choosing a suitable state model for mobile robots, we can use a simple Lyapunov function to achieve parking control, even with feedback noise.

Similarly, in wearable assistive robotics, sensory modality selection plays a vital role in noise suppression. For instance, in an indoor navigation system for the visually impaired [3], tactile and floor-embedded passive RFID tags were deployed instead of continuous range-finding sensors (such as ultrasonic or optical transducers). This architectural choice eliminates false echoes and measurement noise caused by environmental acoustic reflections, ambient humidity, and temperature variations, providing deterministic landmark-based state localization.

While humans and engineering systems use different strategies and techniques to overcome cognitive distraction and measurement noise, the goal is to ensure optimal performance. By designing robust systems to these factors, we can ensure that they perform optimally even in the presence of cognitive distraction and measurement noise.

Crucially, in human-centered assistive interfaces, noise attenuation is not purely an engineering problem. Excessive feedback frequency risks inducing cognitive sensory overload and acoustic distraction for visually impaired users. Mitigating cognitive distraction requires selective, discrete cue generation delivering auditory prompts only at topological decision vertices, thereby aligning engineering signal filtering with the human brain's finite attentional bandwidth.

## 5 Attention Mechanisms in Artificial Intelligence

The interaction between attention mechanisms and artificial intelligence (AI) is an interesting topic that is worth extensive study. Some important approaches in this domain include Transformer Models [4], Bidirectional Encoder Representations from Transformers (BERT) [5], Noise Reduction and Feature Selection, and Sparse Autoencoders [6].

Transformer Models use self-attention mechanisms to focus on relevant parts of input sequences and excel in natural language processing (NLP) tasks. BERT learns contextual representations by attending to surrounding words and has achieved state-of-the-art results on various NLP tasks.

Noise Reduction and Feature Selection are techniques AI models use to filter out irrelevant features during training, akin to selective attention. Sparse Autoencoders learn compact representations by emphasizing essential features, and they have been used for dimensionality reduction and to reconstruct a model through backpropagation.

Therefore, both humans and AI aim for optimal performance by focusing on relevant information. Robustness is also key, as AI models must handle noisy data and adapt to changing conditions. Resource allocation is another important consideration, as AI systems allocate computational resources efficiently, just as our attention capacity is limited.

## 6 Conclusion

Overall, while there are certainly differences between human cognition and engineering systems, there are also important similarities that suggest that there may be common strategies for attenuating and filtering out the effects of noise and distraction. Further research in this area could have significant implications for both cognitive psychology and engineering. In summary, selective attention is a universal principle—whether in our brains, engineering systems, or AI. Understanding these parallels can inspire better designs and enhance performance across domains.